\documentclass[twocolumn,aps,pra,superscriptaddress,amsfonts]{revtex4-1}

\usepackage{amsmath}
\usepackage{amssymb}
\usepackage{graphicx}
\usepackage{bm}
\usepackage{array}
\usepackage{dcolumn}
\usepackage{hepparticles}
\usepackage{heppennames}
\usepackage{hepnicenames}
\usepackage{epstopdf}
\usepackage{tikz}
\usepackage{xcolor}

\newcommand{\ket}[1]{{| {#1} \rangle}}

\begin{document}

\title{Optimized pulsed sideband cooling and enhanced thermometry of trapped ions
}

\author{A.J. Rasmusson$^1$, Marissa D'Onofrio$^1$, Yuanheng Xie$^1$, Jiafeng Cui$^1$, and Philip Richerme}
\affiliation{Indiana University Department of Physics, Bloomington, Indiana 47405, USA}
\affiliation{Indiana University Quantum Science and Engineering Center, Bloomington, Indiana 47405, USA}

\date{\today}

\begin{abstract}
Resolved sideband cooling is a standard technique for cooling trapped ions below \mbox{the Doppler limit} to near their motional ground state. Yet, the most common methods for sideband cooling implicitly rely on low Doppler-cooled temperatures and tightly confined ions, and they cannot be optimized for different experimental conditions. Here we introduce a framework which calculates the fastest possible pulsed sideband cooling sequence for a given number of pulses and set of experimental parameters, and we verify its improvement compared to traditional methods using a trapped $^{171}$Yb$^+$ ion. After extensive cooling, we find that the ion motional distribution is distinctly non-thermal and thus not amenable to standard thermometry techniques. We therefore develop and experimentally validate an improved method to measure ion temperatures after sideband cooling. These techniques will enable more efficient cooling and thermometry within trapped-ion systems, especially those with high initial temperatures or spatially-extended ion wavepackets.
\end{abstract}

\maketitle

\section{Introduction}
\label{sec:intro}
The cooling of mechanical oscillators to near their ground motional state is of fundamental importance to fields as varied as atomic clocks \cite{ludlow2015optical,huntemann2016single,brewer2019al+}, quantum computation and simulation \cite{pino2021demonstration, wright2019benchmarking, blatt2012quantum, monroe2021programmable}, quantum sensing and transduction \cite{biercuk2010ultrasensitive,degen2017quantum,teufel2011sideband}, and even gravitational wave detection \cite{whittle2021approaching}. Particularly for atom-based platforms, Doppler laser cooling provides a fast and straightforward method for reducing the kinetic energy of the system by orders of magnitude to reach the quantum regime \cite{leibfried2003quantum,metcalf1999laser}. Even so, recoil effects during photon emission typically prevent Doppler-cooled systems from achieving their absolute motional ground state, requiring the implementation of sub-Doppler cooling methods \cite{dalibard1989laser,kasevich1992laser,roos2000experimental,ejtemaee20173d}.

For trapped ion experiments, resolved sideband cooling (SBC) is the most popular sub-Doppler cooling technique used to prepare systems near their motional ground state \cite{diedrichlaser1989, marzoli1994laser, monroe1995resolved, king1998cooling}. Its widespread use stems largely from its applicability to most trapped-ion setups, since its effectiveness does not rely on using a specific ion species or trap geometry \cite{wineland1998experimental,chen2020efficient-sideband-cooling}. In practice, SBC allows trapped ions to be initialized in a nearly-pure state of motion, with a typical average harmonic occupation $\bar{n} \lesssim 0.05$ \cite{diedrichlaser1989}. However, SBC is often the longest time component in an experimental cycle by a significant factor \cite{pino2021demonstration}, especially when many motional modes need to be cooled. Although individual addressing can facilitate some speedups in long ion chains \cite{chen2020efficient-sideband-cooling}, to date no general method is known for determining the optimal SBC protocol.

Accurate ion thermometry goes hand-in-hand with near-ground-state cooling techniques such as SBC. Estimating ion temperatures and heating rates are essential characterizations in ion trap experiments \cite{brownnutt2015ion, wineland1998experimental} since they inform the efficacy of cooling protocols and potential sources of noise. Yet, standard methods for measuring $\bar{n}$ near the ground state implicitly assume the motion is well-described by a thermal distribution of harmonic oscillator levels \cite{diedrichlaser1989,monroe1995resolved}. When this assumption is violated, as is the case for Fock states, coherent states, or states following significant SBC \cite{meekhof1996generation,chen2017sympathetic}, more sophisticated thermometry methods must be employed to accurately characterize ion motional temperatures.

Here, we present a framework for calculating the optimal sequence of SBC pulses for near-ground-state cooling, and we develop an improved thermometry technique to more accurately measure $\bar{n}$ following SBC. Our optimal cooling strategy is applicable to any trapped ion experiment using pulsed SBC and flexible enough to incorporate decoherence effects or heating models if desired. Likewise, our method to determine ion temperatures requires only the experimental hardware needed for implementing pulsed SBC. We benchmark both our optimized SBC sequences and our new thermometry technique using a trapped $^{171}$Yb$^+$ ion, finding close experimental agreement with theory predictions as well as significant improvements compared with traditional cooling and thermometry protocols.

The article is structured as follows. Section \ref{sec:sct} reviews the standard theory of pulsed resolved SBC. In Sec. \ref{sec:osp} we recast the pulsed SBC problem into a matrix formalism that allows for efficient numerical optimization of SBC pulse sequences. Section \ref{sec:tscd} introduces a new experimental technique to accurately measure ion temperatures following sub-Doppler cooling, followed by experimental validation in Sec. \ref{sec:et}. We summarize with concluding remarks in Sec. \ref{sec:con}.

\section{Resolved Sideband Cooling Theory}
\label{sec:sct}

When a trapped ion of mass $m$ is confined to a 1D harmonic potential of frequency $\omega$, resolved SBC allows for sub-Doppler cooling of the ion temperature. Prior to the onset of SBC, we assume that the ion has been Doppler cooled using a transition of linewidth $\Gamma$ to the Doppler cooling limit \cite{stenholm1986semiclassical, eschner2003laser}
\begin{equation}
    \label{eqn:dop-limit}
    \bar{n}_i\approx \frac{\Gamma}{2\omega}.
\end{equation}
Following Doppler cooling, the probability of finding the ion in the $n^{\text{th}}$ harmonic oscillator level is well-described by the thermal distribution
\begin{equation}
\label{eqn:thermal-dist}
    p_{\text{th}}(n) = \dfrac{\bar{n}^{n}}{(\bar{n}+1)^{n+1}}
\end{equation}
which is solely parameterized by the average harmonic state of the ion $\bar{n}$.

SBC protocols may be implemented for both optical and hyperfine qubits; here we begin by focusing on the latter. Typically, far-detuned Raman transitions of wavelength $\lambda$ and linewidth $\gamma_{\text{rad}} \ll \omega$ are used to manipulate the electronic and motional states of the ion. When the Raman transition frequency is in resonance with the qubit splitting, it drives a ``carrier" transition between qubit levels $\ket{\downarrow}$ and $\ket{\uparrow}$ at Rabi frequency $\Omega$, with no change to the motional state. Detuning the Raman frequency by integer multiples of the trap secular frequency $\omega$ excites a ``sideband'' transition, coupling spin flips to a change in motional state from $\ket{n}$ to $\ket{n'}$, at Rabi rate \cite{wineland1979laser, wineland1998experimental}
\begin{equation}
\label{eqn:rabi-freq}
    \Omega_{n,n'} = \Omega e^{-\eta^2/2}\sqrt{\dfrac{n_<!}{n_>!}}\eta^{|n-n'|}\mathcal{L}^{|n-n'|}_{n_<}(\eta^2),
\end{equation}
where $n_{<}$ ($n_{>}$) is the lesser (greater) of $n$ and $n'$, 
\begin{equation}
    \mathcal{L}^{(\alpha)}_n(X) = \sum^{n}_{i=0}(-1)^i \binom{n+\alpha}{n - i} \dfrac{X^i}{i!}
\end{equation}
is the generalized Laguerre polynomial, and 
\begin{equation}
    \eta \equiv \Delta k x_0 = 2 \sin(\theta/2) \frac{2\pi}{\lambda} \sqrt{\frac{\hbar}{2 m \omega}}.
\end{equation}
is the Lamb-Dicke parameter for counter-propagating Raman beams which intersect at an angle $\theta$. In this article, we will refer to an $n-n'=1$ transition as a first-order red sideband (RSB) transition and an $n-n'=-1$ transition as a first-order blue sideband (BSB) transition.

SBC of hyperfine qubits is typically characterized by a sequence of discrete RSB pulses interleaved with optical pumping. A traditional pulsed SBC protocol (which we will call the ``classic'' protocol) executes as follows \cite{diedrichlaser1989, monroe1995resolved}. After Doppler cooling to an average harmonic occupation $\bar{n}_i$, and optical pumping to the qubit state $\ket{\downarrow}$, an initial motional level $n_i \gg \bar{n}_i$ is selected as the entry point for SBC. A first-order RSB $\pi$-pulse is then applied for $t = \pi / \Omega_{n_i,n_i-1}$ followed by fast optical pumping, to drive the transition $|\downarrow, n_i\rangle \rightarrow |\downarrow, n_i-1\rangle$. Then another iteration is performed using $t = \pi / \Omega_{n_i-1,n_i-2}$, and so on, until the sequence concludes with a final $t = \pi / \Omega_{1,0}$ pulse. In principle, this protocol sweeps the fraction of population for which $n \leq n_i$ into the motional ground state.

By starting at larger $n_i$ and iterating for more pulses, the classic SBC protocol can theoretically reach the SBC limit of $\bar{n}_{\text{min}} \approx (\gamma_{\text{rad}} /2 \omega)^2 \ll 1$ \cite{wineland1998experimental, neuhauser1978optical, wineland1979laser, wineland1987laser}. In practice, the achievable final $\bar{n}$ may be limited by effects such as imperfect RSB $\pi$-pulses, motional heating, and nearly-infinite RSB $\pi$-times (Sec. \ref{sec:mo}); this is indeed the case for several trapped-ion experiments \cite{che2017efficient-raman,chen2017sympathetic,d2020radial}. Nevertheless, post-SBC temperatures of $\bar{n} \lesssim 0.05$ are routinely achieved with the classic method \cite{diedrichlaser1989,monroe1995resolved}, particularly when the initial state before SBC is in the ``low $\eta$-$\bar{n}_i$ regime": $\eta \ll 1$ and $\bar{n}_i \lesssim 10$.

For optical qubits, continuous SBC is the preferred protocol for achieving near-ground state cooling \cite{roos1999quantum}. In this approach, a RSB is driven continuously on a narrow optical transition while optical pumping is accomplished by spontaneous emission from the excited state. Given the slow decay rate of narrow transitions, spontaneous emission may be enhanced by temporarily coupling the excited state to a dipole-allowed transition. In $^{40}$Ca$^+$, for instance, coupling the quadrupole $D_{5/2}$ qubit level to the dipole-allowed $P_{3/2}$ state can lead to cooling rates of $\dot{\bar{n}} =$ \mbox{5 ms$^{-1}$} when strongly saturating the RSB transition \cite{roos1999quantum}. As we will show in Sec. \ref{sec:osp}, this rate is comparable to the pulsed SBC rate in hyperfine qubits driven by a carrier Rabi frequency $\Omega \approx 2\pi \times 10$ kHz. For our experiments in Sec. \ref{sec:et} we set $\Omega = 2\pi \times 65$ kHz, leading to an initial cooling rate of $\dot{\bar{n}} \approx$ \mbox{30 ms$^{-1}$}.

Continuous SBC has been well-described via detailed theoretical models \cite{marzoli1994laser,eschner2003laser} and validated in experiments \cite{roos1999quantum}. For a given optical pumping rate, the optimum RSB parameters for achieving the lowest final $\bar{n}$ may be estimated from the full set of atomic rate equations \cite{marzoli1994laser}, or determined experimentally by scanning over different values of RSB power and frequency \cite{hempel2014digital}. In contrast, the discreteness of pulsed SBC protocols prevents a similar rate-equation type analysis while greatly expanding the parameter space of possible cooling sequences. For these reasons, finding a pulsed SBC model that allows for efficient determination of optimal sequences has remained elusive to date; we seek to address this open question in the remainder of this article.

\section{Optimized Pulsed SBC Protocols}
\label{sec:osp}
For hyperfine qubits, the intuitive `classic' protocol introduced in Sec. II is not the most \emph{efficient} pulsed SBC method for reducing ion temperatures. Given a chosen $n_i$, which sets the number of pulses, there are no adjustable parameters that may be used to optimize the cooling rate per pulse or per unit time. When starting from small Doppler-cooled $\bar{n}_i$, only a few pulses are needed and the deviation from optimal is small; when $\bar{n}_i$ is large ($\gtrsim 10$), the deviation from optimal widens considerably. If $\bar{n}_i$ is large enough, the classic method will fail to prepare ions in the ground motional state as mentioned previously in Sec. \ref{sec:sct}. 

In this section, we introduce two globally-optimized pulsed SBC protocols: a single-parameter protocol called the ``fixed'' method, and a full-parameter protocol called ``optimal'' method. For a given number of pulses, the optimal method provides the lowest possible $\bar{n}$ after first-order SBC. When $\bar{n}_i$ is large, we show how these protocols can be extended to higher-order SBC to avoid the limitations of first-order cooling. To compute these optimized SBC protocols we must first numerically simulate the complicated interplay between each $\pi-$pulse and its effect on the \emph{entire} harmonic oscillator population $p(n)$. Below, we develop a graph-theoretic description of pulsed SBC to accomplish this task and provide a framework for fast optimization of pulse sequences.

\subsection{Graph-Theoretic Description of Pulsed Sideband Cooling}
\label{sec:gdsc}
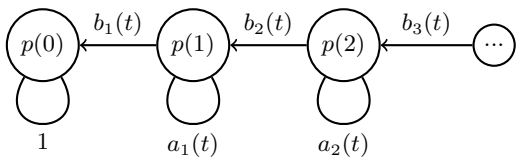
\begin{figure}
    \centering
    \begin{tikzpicture}[node distance={20mm}, thick, main/.style = {draw, circle}] 
    \node[main] (1) {$p(0)$};
    \node[main] (2) [right of=1] {$p(1)$};
    \node[main] (3) [right of=2] {$p(2)$};
    \node[main] (4) [right of=3] {...};
    \draw[->] (2) -- node[midway, above] {$b_1(t)$} (1);
    \draw[->] (3) -- node[midway, above] {$b_2(t)$} (2);
    \draw[->] (4) -- node[midway, above] {$b_3(t)$} (3);
    \draw[-] (1) to [out=240,in=300,looseness=5] node[midway, below] {$1$} (1);
    \draw[-] (2) to [out=240,in=300,looseness=5] node[midway, below] {$a_1(t)$} (2);
    \draw[-] (3) to [out=240,in=300,looseness=5] node[midway, below] {$a_2(t)$} (3);
    \end{tikzpicture}
    \caption{Graph $G$ representing first-order SBC. The set of vertices $V$ is represented by circles and weighted by the current harmonic probability distribution $p(n)$. The set of edges $E$ is represented by lines: loops weighted by $a_n(t)$ and directed edges weighted by $b_n(t)$.}
    \label{fig:graph}
\end{figure}

We embed SBC into a graph $G=(V, E)$ with a set of vertices $V$ and edges $E$. The vertices $V$ represent a truncated set of the harmonic states $n=[0, n_{\text{max}}]$ where $n_{\text{max}} \gg \bar{n}_i$ is well satisfied. Each vertex is weighted by the probability corresponding to its harmonic state $V = \{p(0), p(1),\hdots, p(n_{\text{max}})\}$, as shown in Fig. \ref{fig:graph}. Each vertex has an undirected edge loop weighted by the probability of not cooling: $a_n(t) = \cos^2(\Omega_{n,n-1} t/2)$ in the case of first-order cooling shown in Fig. \ref{fig:graph}. The probability of cooling $b_n(t) = \sin^2(\Omega_{n,n-1} t/2)$ weights a directed edge from the $n$ to $n-1$ vertices. For $m$th-order cooling, the directed edges would connect to their $m$th leftmost neighbor with the associated Rabi frequency $\Omega_{n,n-m}$.

To model one SBC pulse of time $t_0$, all vertex weights take one traversal of their respective edges resulting a new set of vertex weights: $V^{(1)}_{n} = a_n(t_0) V_n^{(0)} + b_{n+1}(t_0) V_{n+1}^{(0)}$. To model $N$ SBC pulses, the graph is traversed $N$ times. In general, each traversal may have its own associated pulse time $\{t_0,t_1,\ldots,t_{N-1}\}$.

\begin{figure}[t!]
    \centering
    \includegraphics[width=8.6cm]{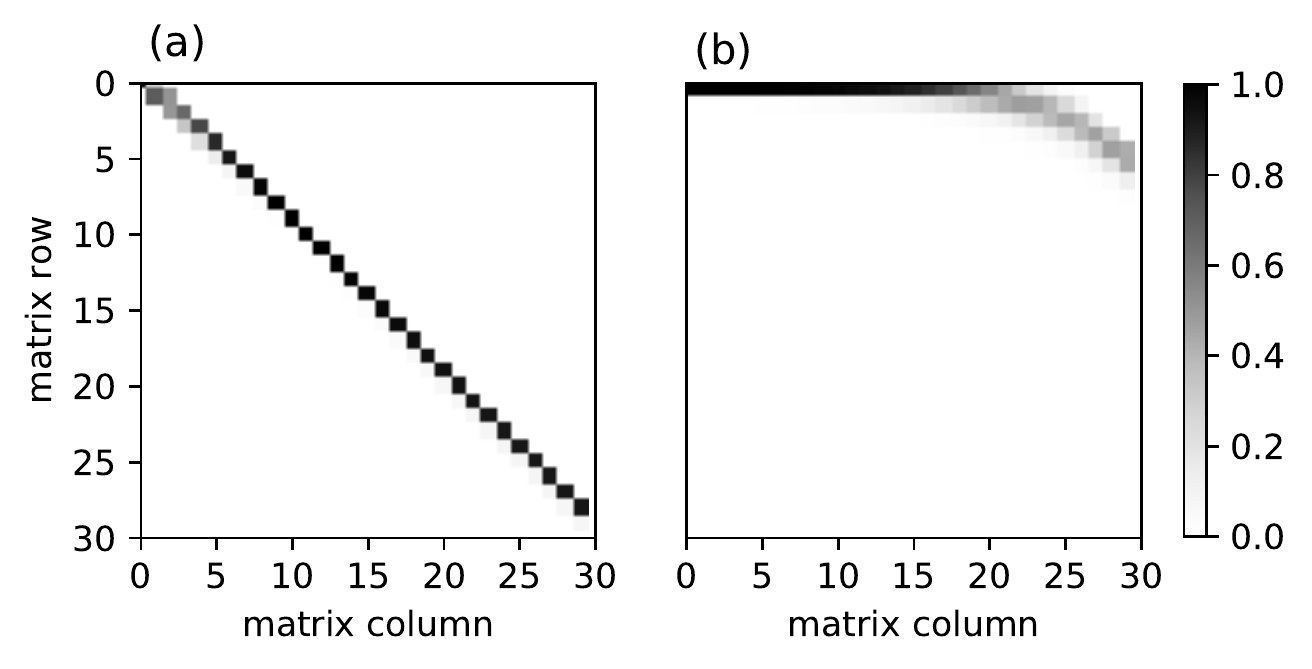}
    \caption{The first $30\times 30$ matrix elements of the weight matrix (Eq. (\ref{eqn:graph-matrix})) are shown graphically for (a) a single pulse and (b) 25 repetitions of the pulse applied in (a).}
    \label{fig:weight-matrices}
\end{figure}

We numerically represent the graph and SBC process as a matrix equation. The initial vertex values map to the vector $\vec{p}_{\text{th}}=\{p_{\text{th}}(0),\hdots,p_{\text{th}}(n_{\text{max}})\}$, where $p_{\text{th}}(n)$ is the initial thermal distribution following Doppler cooling (Eq. (\ref{eqn:thermal-dist})). One traversal of the graph maps to the upper triangular matrix
\begin{equation}
\label{eqn:graph-matrix}
W(t)
=
\begin{pmatrix}
1 & b_1(t) & 0 & \hdots \\
0 & a_1(t) & b_2(t) & \hdots \\
0 & 0 & a_2(t) & \hdots \\
\vdots & \vdots & \vdots &\ddots \\
\end{pmatrix}
\end{equation}
which is shown graphically in Fig \ref{fig:weight-matrices}(a) for $t=1.016\times 2\pi/\Omega$. $W(t)$ acting on $\vec{p}_{\text{th}}$ results in an updated probability vector $\vec{p} = \{p(0), \hdots,p(n_{\text{max}})\}$

\begin{equation}
\begin{pmatrix}
p(0) \\
p(1) \\
p(2) \\
\vdots\\
\end{pmatrix}
=
\begin{pmatrix}
1 & b_1(t) & 0 & \hdots \\
0 & a_1(t) & b_2(t) & \hdots \\
0 & 0 & a_2(t) & \hdots \\
\vdots & \vdots & \vdots &\ddots \\
\end{pmatrix}
\begin{pmatrix}
p_{\text{th}}(0) \\
p_{\text{th}}(1) \\
p_{\text{th}}(2) \\
\vdots \\
\end{pmatrix}.
\end{equation}

To encode the effects of multiple SBC pulses, all individual pulse matrices $W(t)$ are multiplied together: $W(t_{N-1})\ldots W(t_1)W(t_0)$. In the simplest case, when all pulses are of the same duration $t_0$, the SBC interaction is encoded as a matrix power of $W(t_0)$. For example, the final harmonic level occupation after 25 identical pulses can be calculated as $\vec{p} = W^{25}(t_0) \vec{p}_{\text{th}}$, with the low-$n$ matrix elements of $W^{25}$ shown in  Fig. \ref{fig:weight-matrices}(b).


\subsection{Fixed protocol}
\label{sec:so}
Optimized pulse sequences may be efficiently computed within the graph-theoretic framework introduced above. To begin, we consider a single-parameter optimization that we call the ``fixed'' protocol. Each of the SBC pulses is chosen to have the same duration $T_{\text{fixed}} = \{t_0, \ldots, t_0\}$, similar to SBC schemes implemented in some trapped-ion studies \cite{deslauriers2004zero, wan2015efficient, che2017efficient-raman}. Here we explicitly seek to minimize the function 
\begin{equation}
\label{eq:fixedmin}
\bar{n}(t_0) = \sum_{n=0}^{n_{\text{max}}}n\left[W^N(t_0)\vec{p}_{\text{th}}\right]_n
\end{equation}
to find the time $t_0$ which yields the lowest possible $\bar{n}$ given $N$ identical SBC pulses.

The optimal pulse time for the fixed method can be computed quickly since there is only one parameter to optimize for any number of pulses $N$. The most costly step in minimizing Eq. (\ref{eq:fixedmin}) is the calculation of $[W^N(t_0)\vec{p}_\text{th}]_n$ for different $t_0$. However, standard numerical packages, such as python's NumPy module \cite{harris2020array}, can exponentially reduce the number of matrix multiplications needed when computing a power of a matrix through binary decomposition. Assuming $N>3$, a binary decomposition recursively squares the matrix, exponentially increasing the matrix power: $2$, $4$, $8$, and so on. The implementation is adapted to allow for arbitrary matrix powers, with a computation time scaling with $N$ as $\mathcal{O}(\log_2(N))$ and with system size $n_{\text{max}}$ as $\mathcal{O}(n^3_{\text{max}})$.



\subsection{Optimal Protocol}
\label{sec:go}

We now consider the optimal protocol, which is a full-parameter optimization where each pulse time is treated as an independent variable. Given a set of experimental parameters, and restricting---for now---to first-order RSB pulses, the remaining degrees of freedom are the durations of each SBC pulse. The optimal protocol searches the full available parameter space of $N$ distinct pulse times, yielding the lowest possible $\bar{n}$ for any given value of $\eta$, $\bar{n}_i$, $\Omega$, and $N$.

The optimal protocol, using first-order RSBs, executes as follows. First, the initial harmonic populations $\vec{p}_{\text{th}}$ and Rabi frequencies $\Omega_{n,n-1}$ are calculated over a truncated range of harmonic states $[0, n_{\text{max}}]$ ($n_{\text{max}} \gg \bar{n}_i$), based on the experimental parameters $\eta$, $\bar{n}_i$, and $\Omega$.
Next, a gradient descent algorithm is applied to minimize the equation
\begin{equation}
\bar{n}(t_0, t_1, \hdots, t_{N-1}) = \sum_{n=0}^{n_{\text{max}}}n \left[W(t_{N-1}) \hdots W(t_1) W(t_0)\vec{p}_{\text{th}}\right]_{n}
\end{equation}
to find the pulse schedule $T_{\text{optimal}}=\{t_0,t_1,\hdots,t_{N-1}\}$ that gives the lowest average harmonic occupation $\bar{n}(t_0,t_1,\hdots,t_{N-1})$ following $N$ SBC pulses.


Since each pulse time in the pulse schedule $T_{\text{optimal}}$ is an independent variable, computing the optimal $T_{\text{optimal}}$ scales exponentially with the number of pulses. For large $n_\text{max}$ or $N$, this can cause calculations to exceed readily available computational resources. However, we find that careful bounding of the gradient descent minimization can help reduce computation times. For example, using a standard laptop we observe that a 50-pulse SBC optimization takes less than 90 seconds to compute, which is a factor of two faster than for the unbounded case.


\begin{figure}
    \includegraphics[width=8.6cm]{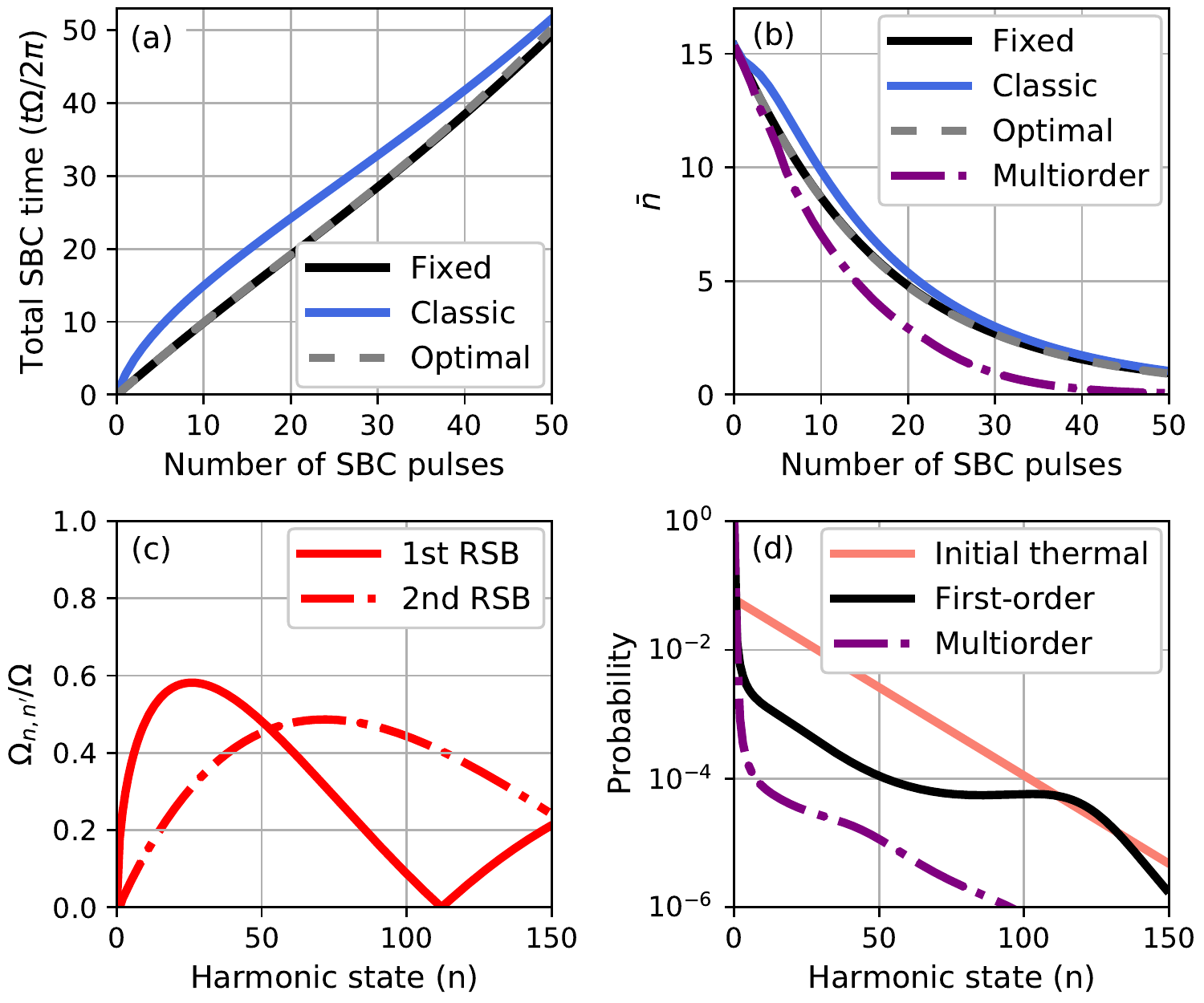}
    \caption{The classic, fixed, optimal, and multiorder protocols are compared for an initial temperature of $\bar{n}_i = 15.36$, and $\eta=0.18$ (see text for definitions). (a) The total sideband cooling time (excluding optical pumping), and (b) the cooled $\bar{n}$ as a function of the number of SBC pulses. (c) Scaled frequencies for the first-order (solid) and second-order (dash-dot) RSB showing the near-zero frequency of the first-order RSB at $n=112$. (d) Initial thermal distribution (solid light red) and distributions after 50 pulses of first-order fixed (solid black) and multiorder fixed (dash-dotted purple).}
    \label{fig:protocol-comparison}
\end{figure}

The predicted performance of the optimal, fixed, and classic protocols are compared in Fig. \ref{fig:protocol-comparison}. Simulations are performed using the parameters $\bar{n}_i=15.36$ and $\eta=0.18$, which are similar to those of our experimental system described in Sec \ref{sec:et}. For fewer than $\sim 50$ SBC pulses, the classic method not only takes the longest absolute time to implement (Fig. \ref{fig:protocol-comparison}(a)), but also yields the highest final $\bar{n}$ (Fig. \ref{fig:protocol-comparison}(b)). In comparison, the fixed (solid black) and optimal (dashed gray) methods perform nearly identically, both in overall cooling time and final ion temperature. For larger $\bar{n}_i$, the classic method drifts further away from optimal, while the fixed method retains its near-optimal behavior. 

\subsection{Multiorder Optimization}
\label{sec:mo}

When outside of the low $\eta$-$\bar{n}_i$ regime, the trapping of harmonic population in high-$n$ states can limit first-order RSB cooling \cite{wan2015efficient, che2017efficient-raman, chen2017sympathetic}. As shown in Fig. \ref{fig:protocol-comparison}(c), the first-order RSB Rabi frequency approaches zero for specific high-$n$ harmonic levels (approximately $n=112$ for our chosen parameters). As a consequence, any initial population $n \gtrsim 112$ will be trapped in these high-$n$ states, even while the remaining population $n \lesssim 112$ is swept towards the ground state. 

This population trapping effect is visible in Fig. \ref{fig:protocol-comparison}(d), which shows the harmonic population distribution following 50 first-order SBC pulses. A significant population near $n=112$ remains uncooled, contributing approximately $0.3$ motional quanta to the final value of $\bar{n}$: an order of magnitude higher than the SBC cooling limit and large compared to what is considered near-ground-state cooling. This effect also explains why the three first-order methods in Fig. \ref{fig:protocol-comparison}(a)-(b) begin to converge at large numbers of pulses: the trapped population contributions to $\bar{n}$ dominate at colder temperatures.



To avoid population trapping at high-$n$, higher-order RSB pulses can be incorporated into the SBC protocol. We refer to this scheme as ``multiorder" cooling. Particularly in experimental regimes where $\eta$ or $\bar{n}_i$ are large, trapped populations may be so significant that multiorder cooling is \textit{required} to achieve near-ground-state temperatures \cite{wan2015efficient, che2017efficient-raman, chen2017sympathetic}. This is because the harmonic levels with near-zero RSB Rabi frequencies shift to smaller $n$ as $\eta$ increases, and because larger fractions of the initial population will be trapped at high-$n$ as $\bar{n}_i$ increases.

Multiorder cooling circumvents population trapping since, for different RSB orders, the Rabi frequencies approach zero at different values of $n$. This is illustrated in Fig. \ref{fig:protocol-comparison}(c), where it can be seen that higher RSB orders exhibit their first zeros at higher values of $n$. This allows for multiorder pulse sequences which first move population from high- to intermediate-$n$, then employ first-order pulses to reach the ground state.

The graph-theoretic framework we introduced in Sec. \ref{sec:gdsc} can easily incorporate higher-order pulses. For an $m^{\text{th}}$ order pulse of time $t$, the probability of not cooling is $a_n(t)=\cos^2(\Omega_{n,n-m} t/2)$ and is mapped to the diagonal of the weight matrix $W(t)$. Likewise, the probability that the $m^{\text{th}}$ order pulse takes $\ket{n}\rightarrow\ket{n-m}$ is $b_n(t)=\sin^2(\Omega_{n,n-m} t/2)$ and is mapped to the $m^{\text{th}}$ upper diagonal of $W(t)$. Both the fixed and optimal protocols may then be calculated for multiorder cooling once the $W(t)$ matrices are constructed.

We simulate and optimize a multiorder fixed protocol with $N_3$ third-order pules, $N_2$ second-order pulses, and $N_1$ first-order pulses fixing the total number of SBC pulses $N = N_1 + N_2 + N_3$ and allowing the pulse time to vary per order $\bar{n}(t_1,t_2,t_3) = \sum_{n=0}^{n_{\text{max}}}n\left[W^{N_1}(t_1) W^{N_2}(t_2) W^{N_3}(t_3) \vec{p}_{\text{th}}\right]_n$. $N_1$, $N_2$, and $N_3$ were selected by brute force optimization of a block sequence (detailed in the next paragraph). Figure \ref{fig:protocol-comparison}(b) shows multiorder cooling (dash-dotted purple) working significantly faster than the optimal first-order method, cooling from $\bar{n}_i = 15.36$ to a final $\bar{n}=0.06$ after only 50 pulses. In addition, the multiorder protocol avoids the high-$n$ population trapping present in the first-order sequences. This can be seen in Fig. \ref{fig:protocol-comparison}(d), where population is much more efficiently transferred from high-$n$ to low-$n$ when multiorder pulses are used.

Multiorder cooling introduces further optimization and experimental challenges. For an $N$ pulse SBC protocol that includes $k_m$ pulses of order $m$, there are a factorial number of permutations $(N!/\prod k_m!)$ in which the pulse orders may be sequenced, and an exponential number of $\{k_m\}$ choices which satisfy $\sum k_m = N$. For small numbers of pulses ($N \lesssim 20$), we used a brute force computation to conclude that a ``block'' sequence is best: all $k_m$ pulses of the same order $m$ stay together in a ``block,'' and higher-order $m$ blocks are applied before lower orders. Under this restriction, the number of possible sequences becomes polynomial in the number of applied orders $m$, scaling as $\mathcal{O}(N^{m-1})$.

In practice, applying pulses with arbitrarily high orders is not experimentally feasible. Transition linewidths narrow for higher orders, making resonant excitation difficult. In addition, transition rates decrease, making pulse times impractically long (Eq. \ref{eqn:rabi-freq}). In our experimental demonstration (Sec. \ref{sec:et}), we reliably address RSB transitions up to 3$^{\text{rd}}$ order. If higher RSB orders are needed, but not possible to apply, alternation between lower orders may still remove trapped population \cite{che2017efficient-raman} at the cost of longer pulse sequences.

\section{Thermometry of Sideband Cooled Distributions}
\label{sec:tscd}
In the quantum regime, full ion thermometry requires knowledge of the probabilities $p(n)$ for occupying each harmonic level $n$, so that the average occupation $\bar{n}=\sum n p(n)$ may be calculated. Given the impracticality of measuring dozens or hundreds of probabilities $p(n)$ to high accuracy, thermometry techniques must make assumptions about the underlying distribution $p(n)$. The most common one is to assume that $p(n)$ is thermal, in which case $\bar{n}$ may be extracted by taking the ratio of first-order RSB and BSB transition probabilities \cite{diedrichlaser1989}. However, Sec. \ref{sec:osp} and Fig. \ref{fig:protocol-comparison}(d) demonstrated that sideband-cooled ions can have dramatically non-thermal distributions $p(n)$, depending on the cooling protocol, the number of RSB orders, and the number of cooling pulses. Thus common ion thermometry methods may give widely inaccurate results following extensive sideband cooling, motivating development of a new approach.

In this section, we begin by outlining two common ion thermometry methods, their underlying assumptions, and the reasons they fail to correctly measure ion temperatures following significant sideband cooling. We then introduce a new technique for ion thermometry which has been specifically tailored to reveal ion temperatures after sideband cooling and depends only on the time-averaged value of RSB transitions.

\subsection{Existing Methods}
Nearly all experiments measuring trapped-ion temperatures deep in the quantum regime follow the approach used in Ref. \cite{diedrichlaser1989}, which we call the ``ratio" method. The ion is first initialized in the state $\ket{\downarrow}$, and the first-order red and blue sidebands are then driven with the same power for the same time. If the ion motional distribution is thermal, then the ratio of RSB to BSB transition probabilities can be related to the average harmonic level occupation $\bar{n}$ (Appendix A):
\begin{align}
\label{eqn:ratio}
    r \equiv \dfrac{P^{\text{RSB}}_{\uparrow}(t)}{P^{\text{BSB}}_{\uparrow}(t)} = \dfrac{\bar{n}}{\bar{n} + 1}.
\end{align}
This ratio $r$ may be experimentally determined by fitting absorption lineshapes to frequency scans over the red and blue sidebands (as in \cite{diedrichlaser1989}), or by driving red and blue sidebands on resonance and taking the ratio of the resulting time series.

The ratio method is powerful due to its direct dependence on $\bar{n}$ and experimental ease. However, the ratio method relies on the assumption of a thermal harmonic distribution which is inherently mismatched to the motional distribution of ions following significant sideband cooling (see Fig. \ref{fig:protocol-comparison}(d)). As we will show in Sec. \ref{sec:et}, this assumption can lead to an order-of-magnitude underestimate of the final $\bar{n}$ after only moderate sideband cooling.

When the underlying motional distribution is known to be non-thermal, alternative thermometry methods may provide a better estimate of $\bar{n}$. One popular method performs a frequency-domain analysis of a BSB Rabi oscillation, using singular value decomposition (SVD) to extract the harmonic level probabilities $p(n)$ \cite{meekhof1996generation}. In this method, a BSB oscillation is described as a matrix of transition probabilities $b_n(t_i) = \sin^2(\Omega_{n,n-1}t_i/2)$ acting on the level probability vector $\vec{p}$ to yield the measured fluorescence at each timestep $t_i$. SVD is then used to pseudo-invert the transition probability matrix and isolate the vector of $p(n)$'s (see Appendix B for more detail). This technique has been successfully implemented to measure $\bar{n}$ for both thermal states as well as coherent states \cite{meekhof1996generation}.

Although SVD is a flexible method for measuring $\bar{n}$ in non-thermal distributions, there are several drawbacks. First, data acquisition can take a long time since long-oscillation time series are necessary to accurately determine as many harmonic state probabilities as possible. This is further compounded by the need to perform many thousands of repetitions to keep quantum projection noise low and avoid potential overfitting during the SVD. Additionally, the output probabilities from SVD have no physical boundary constraints such as $0 \leq p(n) \leq 1$ or $\sum p(n)=1$. This has been found to produce large errors when applied to distributions with many non-negligible probabilities at high harmonic level $n$ \cite{meekhof1996generation}, as is the case for the distributions shown in Fig. \ref{fig:protocol-comparison}(d).




\subsection{Modeling Post-SBC Distributions}
The primary reason that the ratio and SVD methods fail to accurately estimate $\bar{n}$ following SBC is that they are not well-matched to the motional state distributions shown in Fig. \ref{fig:protocol-comparison}(d). After SBC, the largest contributions to $\bar{n}$ are often driven by the residual population remaining at large $n$, which is neglected when using a simple thermal approximation or when focusing on only the low-$n$ populations. Thus, improved modeling of the probability distribution $p(n)$ following SBC is a prerequisite for higher-accuracy estimation of ion temperatures.

To date, the most detailed modelling of post-SBC motional distributions was outlined in \cite{chen2017sympathetic}. Using simulated multi-order SBC pulses, it was found that the harmonic level populations were well-approximated by a double thermal distribution:
\begin{equation}
    \label{eqn:double-thermal}
    p_{\text{double}}(n) = \alpha p_{\text{th}}(n|\bar{n}_l) + \left(1-\alpha \right) p_{\text{th}}(n|\bar{n}_h)
\end{equation}
where $\bar{n}_l$ captures the distribution for low $n$ states, $\bar{n}_h$ captures the distribution for high $n$ states, and the total average occupation is $\bar{n} = \alpha \bar{n}_l + (1-\alpha) \bar{n}_h$. Our numeric simulations of multiorder SBC in Fig. \ref{fig:protocol-comparison}(d) likewise demonstrate that the final state populations are well-described by this double-thermal model. In \cite{chen2017sympathetic}, $\bar{n}$ was experimentally determined by first fitting the simulated distribution to extract $\bar{n}_h$, then fitting the experimental data to Eq. (\ref{eqn:double-thermal}) with $\bar{n}_h$ as a fixed parameter.

Here, we seek to generalize Eq. (\ref{eqn:double-thermal}) and develop a measurement protocol that avoids dependence on numeric simulations. To begin, we propose direct measurement to find the harmonic level populations $p_\text{meas}(n)$ up to $n=k$, where $k > \bar{n}_l$. Using this, we compute the remaining population fraction in all levels $n > k$:
\begin{equation}
\label{eq:pnkexp}
    p_\text{rem}(n > k) = 1 - \sum_{n=0}^k p_{\text{meas}}(n).
\end{equation}
Next, we propose direct measurement of the initial thermal state $\bar{n}_i$ before SBC, which we identify as $\bar{n}_h$ in \mbox{Eq. (\ref{eqn:double-thermal})}. Once again the quantity $p(n>k)$ is calculated, this time for the initial thermal distribution
\begin{equation}
\label{eq:pnkth}
    p_\text{th}(n > k) = \sum_{n=k+1}^\infty \frac{\bar{n}_i^n}{(\bar{n}_i+1)^{n+1}}
\end{equation}
The ratio of Eqs. (\ref{eq:pnkexp}) and (\ref{eq:pnkth}) estimates the fraction of states remaining in an approximate thermal distribution of average occupation $\bar{n}_i$. The final $\bar{n}$ is then estimated as
\begin{equation}
    \label{eqn:fancy-double-thermal}
    \bar{n} \approx \sum_{n=0}^k n p_{\text{meas}}(n) + \frac{p_\text{rem}(n > k)}{p_\text{th}(n > k)}\sum_{n=k+1}^\infty n \frac{\bar{n}_i^n}{(\bar{n}_i+1)^{n+1}}.
\end{equation}

The advantage of Eq. (\ref{eqn:fancy-double-thermal}) is that it leverages the most information available from measurement with no direct dependence on simulation. The only remaining element needed is a robust method to measure the individual probabilities of the low-lying harmonic levels, $p(n \leq k)$. In the following section, we introduce a simple technique that reveals these desired motional state populations.

\subsection{Time-averaged Thermometry}
\label{sec:ta}
We propose a ``time-average" measurement protocol which, when combined with Eq. (\ref{eqn:fancy-double-thermal}), provides a high-accuracy estimate of $\bar{n}$ following SBC. This approach is constructed to measure the individual probabilities of the first few harmonic levels. Suppose a trapped ion is initialized in the state $\ket{\downarrow}$. Then, the expected probability of finding the ion in the $\ket{\uparrow}$ state when driven with an $m^{\text{th}}$ order RSB is given by:
\begin{equation}
\label{eqn:rsb-w-decoherence}
    P_{\uparrow,m}^\text{RSB}(t)=\sum_{n=0}^\infty \frac{1}{2}\left[1-e^{-\gamma t}\cos(\Omega_{n+m,n}t)\right]p(n+m).
\end{equation}
where no assumptions have been made about the probability distribution $p(n)$, and decoherence effects at rate $\gamma$ have been included for generality.

The running time average of Eq. (\ref{eqn:rsb-w-decoherence}) is
\begin{align}
    \bar{P}^{\text{RSB}}_{\uparrow,m}(t) & =\frac{1}{t}\int_0^t P_{\uparrow,m}^\text{RSB}(t') dt' \nonumber \\
    & = \frac{1}{2} \sum_{n=0}^\infty p(n+m)\left[1 - \frac{\gamma}{\left(\Omega^2_{n+m,n} + \gamma^2 \right) t} \right. \nonumber \\
    & \left. + \frac{e^{-\gamma t} (\gamma\cos(\Omega_{n+m,n}t') - \Omega_{n+m,n} \sin(\Omega_{n+m,n} t))}{\left(\Omega^2_{n+m,n} + \gamma^2 \right) t}\right].
\end{align}
We observe that for long times ($t\gg 1/(\Omega^2_{n+m,n} + \gamma^2)$), the time average converges to a partial sum of motional state probabilities
\begin{equation}
\label{eq:pbar_rsb}
    \bar{P}^{\text{RSB}}_{\uparrow,m}(t) \approx \frac{1}{2} \sum_{n=0}^\infty p(n+m).
\end{equation}

To extract the individual harmonic probabilities, consider driving with a first-order RSB:
\begin{align}
   \bar{P}^{\text{RSB}}_{\uparrow,1}(t) & \approx \frac{1}{2} \sum_{n=0}^\infty p(n+1) \nonumber \\
   & \approx \dfrac{1}{2}\left[1 - p(0)\right]
\end{align}
from which $p(0)$ can be directly estimated
\begin{equation}
\label{eqn:p0}
    p(0) \approx 1 - 2 \bar{P}^{\text{RSB}}_{\uparrow,1}(t).
\end{equation}
Higher harmonic state probabilities may then be estimated by driving with sequentially higher-order RSBs and applying the recursion relation 
\begin{equation}
\label{eqn:recursive}
    p(m-1)\approx 2(\bar{P}_{\uparrow,m-1}-\bar{P}_{\uparrow,m}).
\end{equation}

This time-average approach provides an efficient and robust method for extracting motional state populations. Compared with existing methods, relatively few points are needed to determine the time average of the RSB oscillation. Although these points should be taken at long times (relative to the RSB Rabi frequency), we note that Eq. (\ref{eq:pbar_rsb}) does not depend on the decoherence rate $\gamma$, and indeed converges \emph{faster} when decoherence is included. Rather, we anticipate that the largest errors in time-average measurements will arise from real-time changes in $p(n)$ driven by motional heating. Such trap heating effects have been comprehensively studied \cite{brownnutt2015ion} and can be incorporated into the motional state analysis if needed.

\section{Experimental Thermometry}
\label{sec:et}
In this section, we experimentally demonstrate the effectiveness of our time-averaged thermometry method. We begin by measuring the temperature of a trapped ion following Doppler cooling and comparing the time-average method to several existing techniques. We then repeat our measurements and comparisons using an optimized sideband cooling sequence from Sec. \ref{sec:osp}, finding that the time-average method most closely agrees with theory predictions. 

Thermometry experiments are performed on a single ${}^{171}\text{Yb}^+$ ion confined in a linear Paul trap with axial frequency $\omega_z=2\pi \times 0.670 \pm 0.008$ MHz. In our setup, the Lamb-Dicke parameter $\eta=0.18 \pm 0.01$, the Rabi carrier frequency $\Omega=2\pi \times 64.9 \pm0.5$ kHz, and the optical pumping time is $5$ $\mu$s. Doppler cooling is performed with 369.5 nm light along the ${}^2S_{1/2} |F = 0\rangle \rightarrow {}^2P_{1/2} |F = 1\rangle$ and ${}^2S_{1/2} |F = 1\rangle \rightarrow {}^2P_{1/2} |F = 0\rangle$ transitions (linewidth $\Gamma = 2\pi\times 19.6$ MHz), while red and blue sideband transitions are performed with far-detuned Raman beams at 355 nm. After each experiment, the qubit state is determined by irradiating the ion with 369.5 nm light resonant with the ${}^2S_{1/2} |F = 1\rangle \rightarrow {}^2P_{1/2} |F = 0\rangle$ transition and capturing the spin-dependent fluorescence on a photomultiplier tube.

\subsection{Thermal Distribution}
When an ion is cooled to its Doppler-limited temperature, the motional state is well-characterized by a thermal distribution (Eq. (\ref{eqn:thermal-dist})). Given our axial trap frequency, this temperature corresponds to an average harmonic occupation $\bar{n}_{\text{Dop}} = 14.6 \pm0.2$ (Eq. (\ref{eqn:dop-limit})). We take this value as the theoretical prediction, against which we compare several different methods for trapped-ion thermometry.
\begin{figure}
    \centering
    \includegraphics[width=8.6cm]{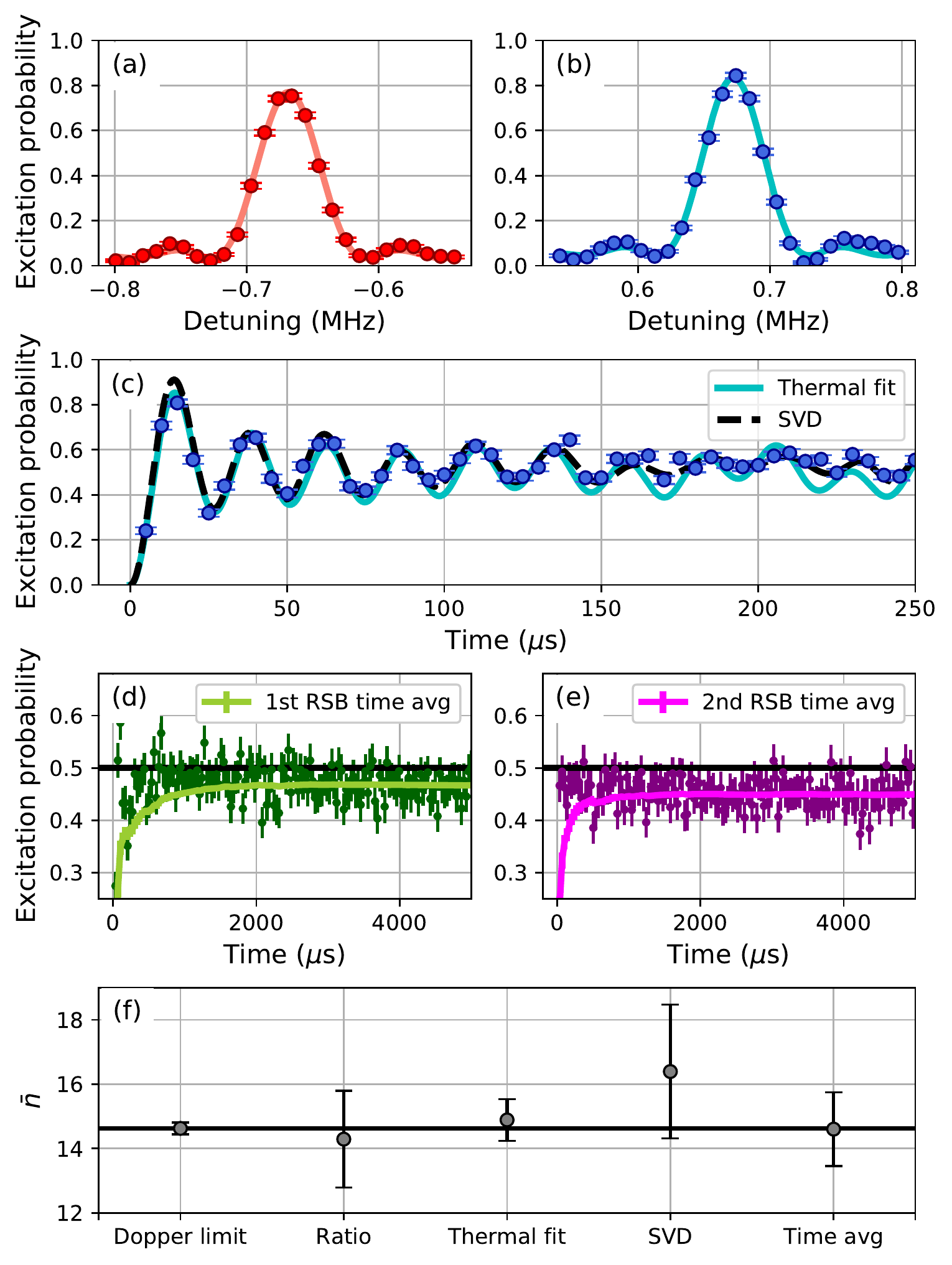}
    \caption{Thermometry comparisons of thermally-distributed ion motional states. (a) and (b) are red and blue sideband frequency scans used to determine $\bar{n}$ from the ratio method. (c) shows BSB Rabi oscillation data (blue points) fit by both a thermally-weighted Rabi oscillation function (solid blue) and a SVD analysis (dashed black). (d) and (e) are long Rabi oscillations of the first- and second-order RSBs, respectively, with their running time-average values shown as solid lines. (f) compares these different thermometry methods against the calculated Doppler cooling limit of $\bar{n}_{\text{Dop}}=14.6$.}
    \label{fig:thermal-comparison}
\end{figure}

We begin by using the ratio method to estimate the Doppler-cooled ion temperature. Figs. \ref{fig:thermal-comparison}(a)-(b) show frequency scans over the red and blue sidebands, respectively, with error bars smaller than the size of the markers. Sinc squared functions are fit to the data with excellent agreement and shown as solid lines. Taking the ratio of the RSB and BSB transition strengths (Eq. (\ref{eqn:ratio})) yields $\bar{n}_{\text{ratio}}=14.3 \pm1.5$, in good agreement with the Doppler-limited prediction. 

Two additional estimates of the Doppler-limited temperature may be extracted by driving a first-order BSB oscillation. In the first method, the data is fit to a thermally-weighted Rabi oscillation $P^{\text{BSB}}_{\uparrow}(t) = \sum^{800}_{n=0}p_{\text{th}}(n)\sin^2(\Omega_{n,n+1}t/2)$, shown as the solid light blue curve in Fig. \ref{fig:thermal-comparison}(c). This single-parameter fit finds an estimated $\bar{n}_{\text{thermal fit}} = 14.9\pm0.7$. Using the same BSB data set, we also employ the SVD method to estimate $\bar{n}_{\text{SVD}}=16.4\pm2.1$. In Fig. \ref{fig:thermal-comparison}(c), the dashed black curve is calculated by weighting a BSB oscillation function $P^{\text{BSB}}_{\uparrow}(t) = \sum^{n_{\text{SVD}}}_{n=0}p_{\text{SVD}}(n)\sin^2(\Omega_{n,n+1}t/2)$ with the SVD-computed probabilities $p_\text{SVD}(n)$. 

Finally, the first (dark green) and second (dark purple) RSBs are driven over a long period of time, with their respective running time averages (light green and light purple) shown in Figs. \ref{fig:thermal-comparison}(d)-(e). We take an excess of data points in our demonstration to confirm the accuracy of this new technique, though we note that only $\sim20$ data points at long times are needed to find the same $\bar{n}$ to within $5\%$.
From the first-order RSB time average in Fig. \ref{fig:thermal-comparison}(d), we estimate $p(0)$ using Eq. (\ref{eqn:p0}). Using the second-order RSB time average in Fig. \ref{fig:thermal-comparison}(e) and the value for $p(0)$, $p(1)$ may be obtained from Eq. (\ref{eqn:recursive}). Finally, fitting $p(0)$ and $p(1)$ to a thermal distribution yields $\bar{n}_{\text{time avg}} = 14.6 \pm 1.2$. 

All extracted values of $\bar{n}$ are compared to the Doppler-limited prediction in Fig. \ref{fig:thermal-comparison}(f). We conclude that all approaches studied here are viable methods for extracting the average harmonic occupation $\bar{n}$ when applied to thermal distributions. In the following subsection, we will re-apply these measurement techniques to sideband cooled ions, whose motional distributions are predicted to be significantly non-thermal.

\subsection{Sideband Cooled Distribution}


In this set of experiments, the ion is initially cooled to the Doppler limit of $\bar{n} = 14.6$, then further cooled using 25 first-order fixed SBC pulses (see Sec. \ref{sec:so}). As shown in Fig. \ref{fig:protocol-comparison}(b), this small number of pulses cannot reach the ground state using any SBC protocol when starting from such a large initial $\bar{n}$. Nevertheless, we will show that 25 SBC pulses is already sufficient to induce large discrepancies between different thermometry techniques.

\begin{figure}
    \centering
    \includegraphics[width=8cm]{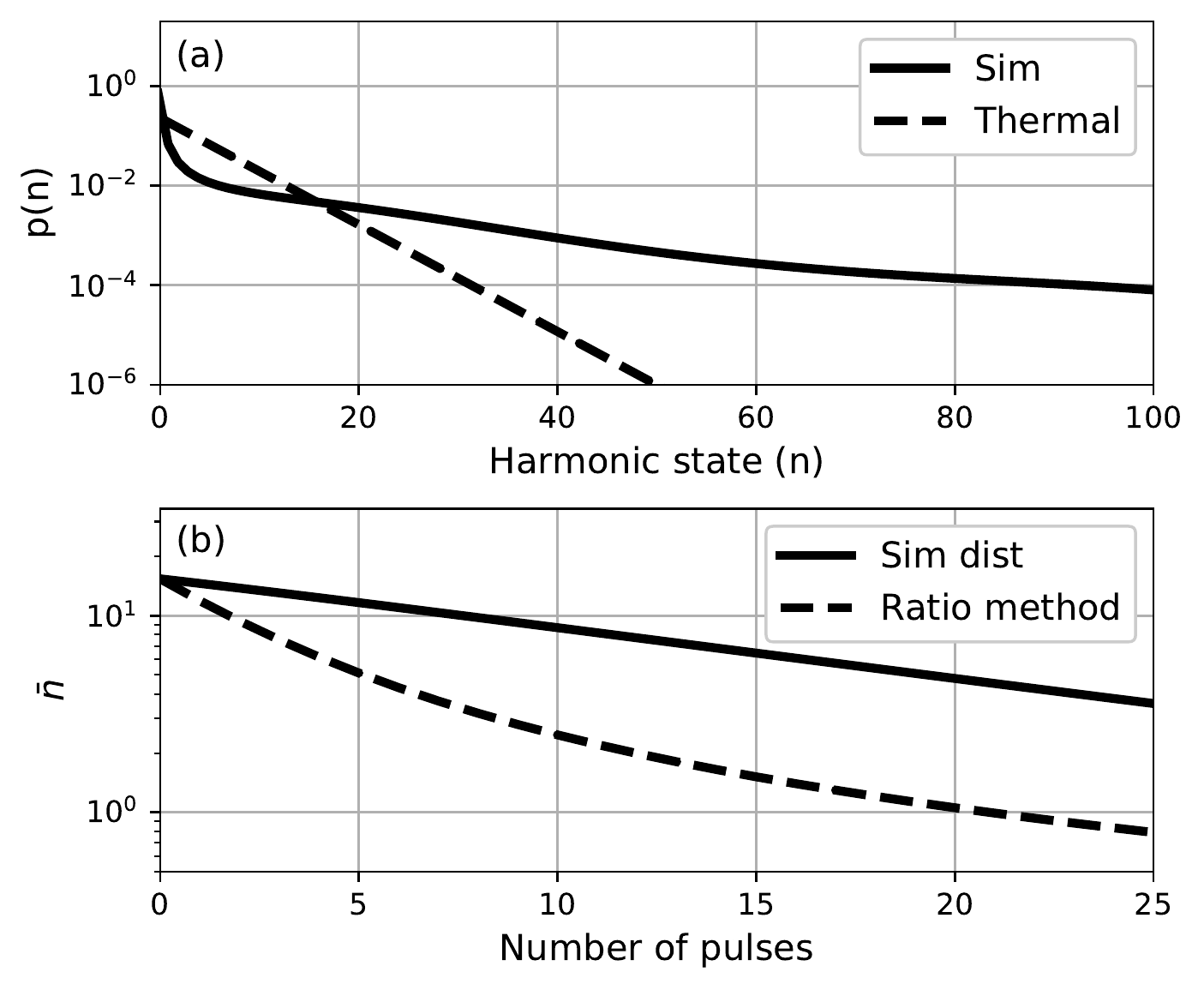}
    \caption{(a) Simulated motional state distribution after 25 first-order fixed pulses (solid), and a thermal distribution with the same $\bar{n}$ (dashed). (b) For any number of SBC pulses, the estimated $\bar{n}$ from the ratio method (dashed) is predicted to significantly underestimate the true $\bar{n}$ as calculated from the simulated distribution (solid).}
    \label{fig:ratio-method-comparison}
\end{figure}

The inherent nonthermal distribution of the sideband cooled ion is predicted to cause a significant bias in the ratio method's estimation of $\bar{n}$. Fig. \ref{fig:ratio-method-comparison} illustrates this point for the given experimental parameters. In Fig. \ref{fig:ratio-method-comparison}(a), a simulated distribution after 25 first-order fixed SBC pulses (solid) is compared to a thermal distribution with the same $\bar{n}$ (dashed). The wide discrepancy indicates that a thermal state is a poor approximation for the post-SBC distribution. 

To quantify the potential error in assuming a thermal distribution, Fig. \ref{fig:ratio-method-comparison}(b) compares the $\bar{n}$ of the simulated distribution (solid) to the predicted result from the ratio method (dashed). The ratio method drastically underestimates $\bar{n}$ after just a few pulses, with almost a full order of magnitude difference by 25 pulses. We caution that when ratio-method thermometry is applied after significant SBC, it may result in misleadingly low estimates of ion temperatures and motional heating rates.

\begin{figure}
    \centering
    \includegraphics[width=8.6cm]{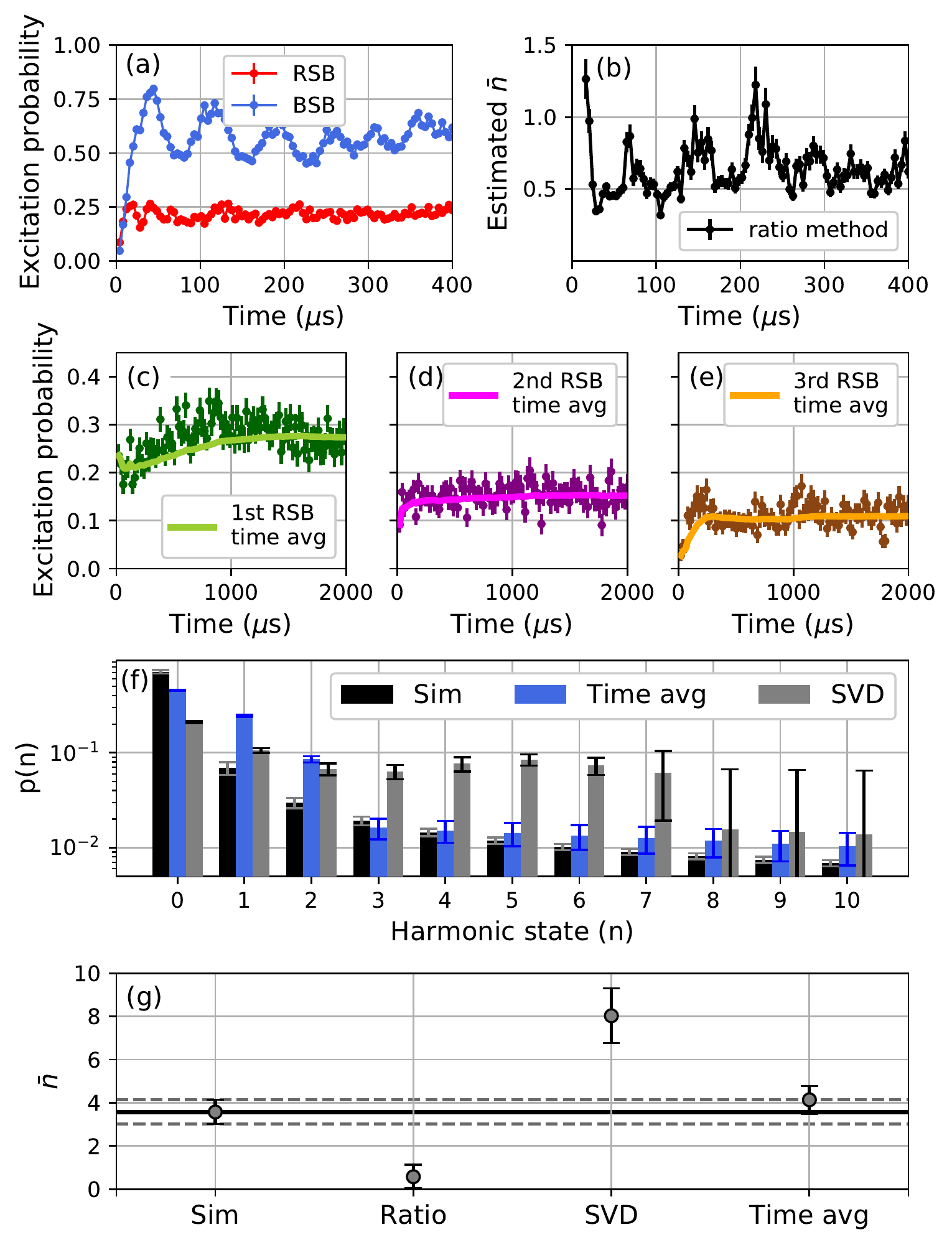}
    \caption{Thermometry comparisons of a sideband cooled ion. (a) the measured first-order RSB and BSB time series. Points are connected to guide the eye. (b) $\bar{n}$ estimation at each time point using the ratio method (excluding the first few time steps). (c)-(e) long-time Rabi oscillations for the first, second, and third RSBs, respectively, with their running time averages drawn as solid lines. (f) population distributions as estimated by numeric simulation (black), time-averaged method (blue), and SVD (gray). (g) $\bar{n}$ measurements from the ratio method, SVD, and the time-average method are compared to a numeric simulation of SBC. Only the time-average method closely estimates $\bar{n}_{\text{sim}}$.}
    \label{fig:nonthermal-comparison}
\end{figure}

Following SBC, we show the Rabi oscillations of first-order red and blue sidebands in Fig. \ref{fig:nonthermal-comparison}(a). The data points are connected (not fitted) to guide the eye, and errors at each point are the size of the marker. Under the assumptions of the ratio method, the ratio of the RSB to BSB at any point in the time provides a valid estimate of $\bar{n}$. We have calculated this ratio for all points in Fig. \ref{fig:nonthermal-comparison}(a), and have plotted the corresponding $\bar{n}$ in Fig. \ref{fig:nonthermal-comparison}(b).

For thermal distributions, as assumed by the ratio method, $\bar{n}$ should be constant at all times. In Fig. \ref{fig:nonthermal-comparison}(b), the substantial differences in extracted $\bar{n}$ with time provide experimental evidence that the underlying state distribution is nonthermal. To estimate $\bar{n}$ in Fig. \ref{fig:nonthermal-comparison}(b), we average over the varying $\bar{n}$ to find $\bar{n}_{\text{ratio}}= 0.58 \pm 0.56$. This value is a drastic \emph{underestimate} of the predicted value $\bar{n}_{\text{sim}} = 3.57 \pm 0.58$, by almost a full order of magnitude. Furthermore, the simulated $\bar{n}$ does not account for ion heating or noise effects, which if included would make the discrepancy even larger.

Next, we applied a SVD analysis to the first-order BSB in Fig. \ref{fig:nonthermal-comparison}(a). Since the tail of the SBC distribution is predicted to be long, we chose the length of the level probability vector $\vec{p}$ to maximize the number of physically constrained probabilities, $0 \leq p(n) \leq 1$. Nevertheless, the BSB time-series data remained poorly fit for any length of $\vec{p}$, and the most accurate SVD result ($\bar{n}{_\text{SVD}} = 8.0 \pm 1.3$) still significantly disagrees with the simulated average harmonic occupation. 


Lastly, we apply our time-average measurement technique to a sideband cooled ion. We begin by driving the the first (dark green), second (dark purple), and third (dark orange) RSBs for a long time period, as shown in Figs. \ref{fig:nonthermal-comparison}(c)-(e). Following the time average procedure outlined in Sec. \ref{sec:ta}, $p(0)$, $p(1)$, $p(2)$, and $p(n>2)$ are estimated from the measured time averages. Substituting these probabilities into Eq. (\ref{eqn:fancy-double-thermal}) results in a measured $\bar{n}_{\text{time avg}} = 4.1 \pm 0.7.$

The estimated level distributions from the simulation, time average method, and SVD method are compared in Fig. \ref{fig:nonthermal-comparison}(f). The numerically simulated distribution (black) follows a monotonic decrease in population for increasing $n$. The time average method (blue) finds similar monotonic behavior, with a relative excess of population in the $n=1$ and $n=2$ levels which we attribute to ion motional heating out of the $n=0$ state \cite{brownnutt2015ion}. In contrast, the distribution estimated by the SVD method (gray) is non-monotonic and exhibits a steep drop-off in population between $n=7$ and $n=8$, suggesting unphysical behavior which cannot be explained by standard heating models \cite{brownnutt2015ion}. Of all the considered thermometry techniques, the time average method best matches the simulated level distributions, 
and it is the only method that does not significantly disagree with the simulated prediction $\bar{n}_{\text{sim}}$ (Fig. \ref{fig:nonthermal-comparison}(g)).



\section{Conclusion}
\label{sec:con}
Sideband cooling has been a popular and powerful technique for the near ground-state preparation of trapped ions. Yet, historical approaches to SBC can be made more efficient, and the measurement of cooled ion temperatures can be performed with less error. In this work, we have shown how to calculate the optimal pulsed SBC protocol for any experimental setup characterized by a cooling laser geometry and wavelength, an ion wavepacket width (which depends upon the ion mass and trap frequency), and an initial ion temperature (which depends on the trap frequency and atomic linewidth). We have additionally argued that careful understanding of the expected state distributions is a necessary precondition for accurate thermometry.

Our efficient numeric simulations and optimizations were enabled by expressing pulsed SBC within a graph-theoretic framework. This approach is powerful for optimizing SBC pulse sequences, and is particularly important in regimes with high Doppler-limited initial temperatures $\bar{n}_i$, or extended ion wavepackets (which correspond to a large $\eta$). We observe that repeated SBC pulses with a single optimized time perform nearly-identically to fully-optimized pulse sequences, while traditional protocols were the least efficient per pulse and per unit time. We have likewise introduced a new thermometry technique which more closely models the state distribution after SBC, and experimentally validated its performance. In contrast, we observe that the most common measurement technique can severely underestimate ion temperatures if extensive SBC is performed.

In future work, we anticipate that the graph representation of pulsed SBC may be expanded to include noise models for ion heating, decoherence, off-resonant couplings, and effects of rf-driven micromotion. Such additions could be smoothly incorporated into the matrix formalism and would allow for further SBC optimization in the face of realistic experimental imperfections. Extending to multiple ions and multiple modes is another natural direction that fits nicely within the matrix representation of pulsed SBC. 

Finally, the time-average technique can open new possibilities for improved thermometry. With this method, for instance, it should be possible to probe the time-dependent population dynamics of trapped-ion motional states and observe how the harmonic level distribution changes in response to external noise sources. Such experiments would provide an additional set of characterizations which may help elucidate mechanisms responsible for anomalous ion heating.

\begin{acknowledgments}
This work was supported by the U.S. Department of Energy, Office of Science, Basic Energy Sciences, under Award $\#$DE-SC0020343. The IU Quantum Science and Engineering Center is supported by the Office of the IU Bloomington Vice Provost for Research through its Emerging Areas of Research program.
\end{acknowledgments}

\bibliographystyle{prsty}
\bibliography{main}{}

\begin{appendix}
\section{Ratio Thermometry}

The ratio method \cite{diedrichlaser1989} estimates the average harmonic state $\bar{n}$ of a thermal distribution $p_{\text{th}}(n) = \bar{n}^n / (\bar{n}+1)^{n+1}$ by using the unique property $p_{\text{th}}(n+1) = p_{\text{th}}(n) \bar{n} / (\bar{n} + 1)$.

Given a RSB Rabi oscillation
\begin{align}
    P^{\text{RSB}}_{\uparrow}(t) & = \sum^{\infty}_{n=1} p_{\text{th}}(n) \sin^2\left(\dfrac{\Omega_{n,n-1}t}{2}\right) \nonumber \\
    & = \dfrac{\bar{n}}{\bar{n}+1}\sum^{\infty}_{n=0} p_{\text{th}}(n) \sin^2\left(\dfrac{\Omega_{n+1,n}t}{2}\right)
\end{align}

and a BSB Rabi oscillation

\begin{align}
    P^{\text{BSB}}_{\uparrow}(t) = \sum^{\infty}_{n=0} p_{\text{th}}(n) \sin^2\left(\dfrac{\Omega_{n+1,n}t}{2}\right)
\end{align}

their ratio is a function of $\bar{n}$ for any time $t$ or frequency detuning

\begin{align}
\label{eqn:ratio}
    r \equiv \dfrac{P^{\text{RSB}}_{\uparrow}(t)}{P^{\text{BSB}}_{\uparrow}(t)} = \dfrac{\bar{n}}{\bar{n} + 1}.
\end{align}

We note that in the presence of decoherence, as introduced in Sec. \ref{sec:ta}, the RSB and BSB transition probabilities may be written

\begin{equation}
    P^{\text{RSB}}_{\uparrow}(t)  = 
    \dfrac{\bar{n}}{\bar{n}+1}\sum^{\infty}_{n=0} p_{\text{th}}(n) \frac{1-e^{-\gamma t}\cos\left(\Omega_{n,n+1}t\right)}{2}
    \end{equation}
    
\begin{equation}
        P^{\text{BSB}}_{\uparrow}(t)  = 
    \sum^{\infty}_{n=0} p_{\text{th}}(n) \frac{1-e^{-\gamma t}\cos\left(\Omega_{n,n+1}t\right)}{2}
\end{equation}
Thus, under this standard model of decoherence, the ratio of RSB to BSB transition probabilities remains identical to the decoherence-free case, $r=\bar{n}/(\bar{n}+1)$.

\bigskip

\section{SVD Thermometry}

The SVD method \cite{meekhof1996generation} is a frequency-domain analysis of a RSB or BSB Rabi oscillation. In this method, $\Omega_{n,n^{\prime}}$ is independently calculated, and its contribution to the overall Rabi oscillation is constructed into a rectangular matrix (dimension $M \times N$) with $M$ time steps taken in the experiment, considering $N$ harmonic states of interest, and elements $b_n(t) = \sin^2(\Omega_{n,n-1}t/2)$. This matrix acts on the harmonic distribution vector ($N \times 1$) to produce a vector representing the measured fluorescence at each experimental time step. For example, a BSB oscillation would be constructed as follows

\begin{equation}
\label{eqn:svd}
\begin{pmatrix}
b_1(t_0) & b_2(t_0) & \hdots \\
b_1(t_1) & b_2(t_1) & \hdots \\
b_1(t_2) & b_2(t_2) & \hdots \\
\vdots & \vdots &\ddots \\
\end{pmatrix}
\begin{pmatrix}
p(0) & \\
p(1) & \\
p(2) & \\
\vdots & \\
\end{pmatrix}
=
\begin{pmatrix}
P^{\text{BSB}}_{\uparrow}(t_0) & \\
P^{\text{BSB}}_{\uparrow}(t_1) & \\
P^{\text{BSB}}_{\uparrow}(t_2) & \\
\vdots & \\
\end{pmatrix}.
\end{equation}

Using singular value decomposition (SVD), the rectangular matrix is pseudo-inverted to solve for the harmonic distribution vector. Once this vector of $p(n)$ is known, the average occupation is found by calculating $\bar{n}=\sum np(n)$.

\end{appendix}
\end{document}